\documentclass{PoS}
\usepackage{amsmath,amssymb,amsfonts}

\title{Scale setting via the $\Omega$ baryon mass}
\ShortTitle{Scale setting via the $\Omega$ baryon mass}

\author{Stefano Capitani$^1$,
        Michele Della Morte$^{1,2}$,
        \speaker{Georg von Hippel}$^1$,
        Bastian Knippschild$^1$,
        Hartmut Wittig$^{1,2}$ \\
        $^1$ Institut f\"ur Kernphysik, Johannes-Gutenberg-Universit\"at Mainz,
        55099 Mainz, Germany\\
        $^2$ Helmholtz-Institut Mainz, University of Mainz, 55099 Mainz, Germany
        \email{hippel@kph.uni-mainz.de}}

\abstract{We present the first results of an ongoing effort to determine the
          lattice scale on the $N_f=2$ CLS lattice ensembles via the mass of the
          $\Omega$ baryon.
          Results from different methods are compared, and various sources of
          systematic uncertainty are discussed.
          \vskip2cm\hfill\includegraphics[width=2cm,keepaspectratio=]{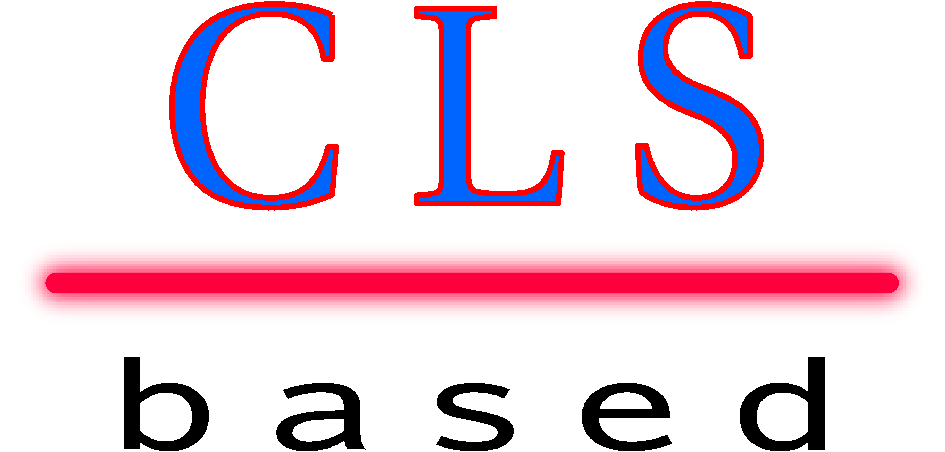}}

\FullConference{XXIX International Symposium on Lattice Field Theory \\
		July 11 -- 16 2011\\
		Squaw Valley, Lake Tahoe, California}

\begin{document}

%% Shorthands
\def\ks{\ensuremath{\kappa_{\rm strange}}}

\section{Introduction}

In order to give absolute values for dimensionful quantities
(including masses and decay constants) from a lattice simulation,
we need to know the lattice spacing $a$.

Since no unique prescription for scale-setting exists, there is a certain
ambiguity arising from the choice of the dimensionful quantity used to
set the scale, which however is at least formally small since it is
of higher order in the lattice spacing. Of more practical importance are
the uncertainties that arise from an inaccurate knowledge of the scale
caused by statistical and systematic errors on the relevant quantity.

Here, we present the result of our effort to set the scale via the mass
of the $\Omega$ baryon
\cite{Brandt:2010ed}.
The $\Omega$ mass has a number of advantages over
other quantities, in that the $\Omega$ is stable in QCD
(as opposed to e.g. the $K^*$ meson),
and that its mass is only very weakly dependent on the light quark mass
(as opposed to e.g. the nucleon).
In addition, all calculations necessary to determine the scale from the
$\Omega$ mass can be done in the fully relativistic theory (as opposed to
e.g. a scale determination from the $\Upsilon$ 2S-1S splitting), and no
renormalization constant is needed for $m_\Omega$ (as opposed to e.g. $f_K$).
Possible disadvantages of using the $\Omega$ baryon include the less favourable
signal-to-noise ratio for baryons, which may render the extraction of a precise
value for $am_\Omega$ difficult. In an $N_{\rm f}=2$ simulation, an additional
potential downside of the $\Omega$ baryon is the quenching of the strange quark,
which may introduce additional (presumably small) systematic effects.

The $\Omega$ baryon thus provides a way to set the scale that reduces many
sources of systematic error in exchange for an increase in statistical error.
To the extent that this increased error can be beaten by performing more
measurements, this is a rather favourable situation.

\subsection{The CLS ensembles}

Coordinated Lattice Simulations (CLS) is a consortium designed to pool
the human and computer resources of several teams in Europe
interested in lattice QCD. CLS member teams are located at CERN, in Germany
(Berlin, DESY/Zeuthen, Mainz, Wuppertal), Italy (Rome, Milan)
and Spain (Madrid, Valencia).
All CLS simulations use either M. L\"uscher's implementation of the
DD-HMC algorithm
\cite{Luscher:2005rx}
or the MP-HMC algorithm by M. Marinkovic and S. Schaefer
\cite{Marinkovic:2010eg}
to efficiently simulate \mbox{$N_f=2$} Wilson QCD with non-perturbative
O($a$) improvement on a variety of computer architectures ranging from
PC clusters to the BlueGene/P at NIC/Forschungszentrum J\"ulich.

The ensembles used in the present analysis are listed in table
\ref{tab:cls}.
\begin{table}[b]
\begin{center}
\begin{tabular}{lllll}\hline\hline
   & $\beta$ & $T\times L^3$ & $\kappa$ & $N_{\rm cfg}$ \\

\hline
A3 & 5.2 & $64\times32^3$ & 0.13580 & 261 \\
A4 &     &                & 0.13590 & 371 \\
A5a&     &                & 0.13594 & 201 \\
\hline
E5c& 5.3 & $64\times32^3$ & 0.13625 & 112 \\
F6 &     & $96\times48^3$ & 0.13635 & 192 \\
F7a&     &                & 0.13638 & 249 \\
\hline
N3 & 5.5 & $96\times48^3$ & 0.13640 & 149 \\
N4 &     &                & 0.13650 & 142 \\
N5 &     &                & 0.13660 & 236 \\
O7 &     & $128\times64^3$& 0.13671 & 78  \\
\hline\hline
\end{tabular}
\end{center}
\caption{CLS ensembles used in this analysis. $N_{\rm cfg}$ is the number
         of configurations used for the calculation of hadron masses in this
         study, not the total number generated.}
\label{tab:cls}
\end{table}
\section{Methods}

\subsection{Measurements}

We measure the two-point correlators for mesons and baryons using
Gaussian-smeared sources
\cite{wavef:wupp1}
with HYP-smeared links
\cite{HYP}
in the covariant Laplacian to suppress excited state contributions.
The parameters of the smearing are fixed at each $\beta$
so as to correspond to a smearing radius of approximately $0.4\textrm{ fm}$.
Multiple point sources per configuration are used in order to improve
statistics. For the vector mesons and baryons, we also average over
multiple polarizations to further reduce statistical errors.

\begin{figure}
\begin{center}
\includegraphics[width=0.32\textwidth,keepaspectratio=]{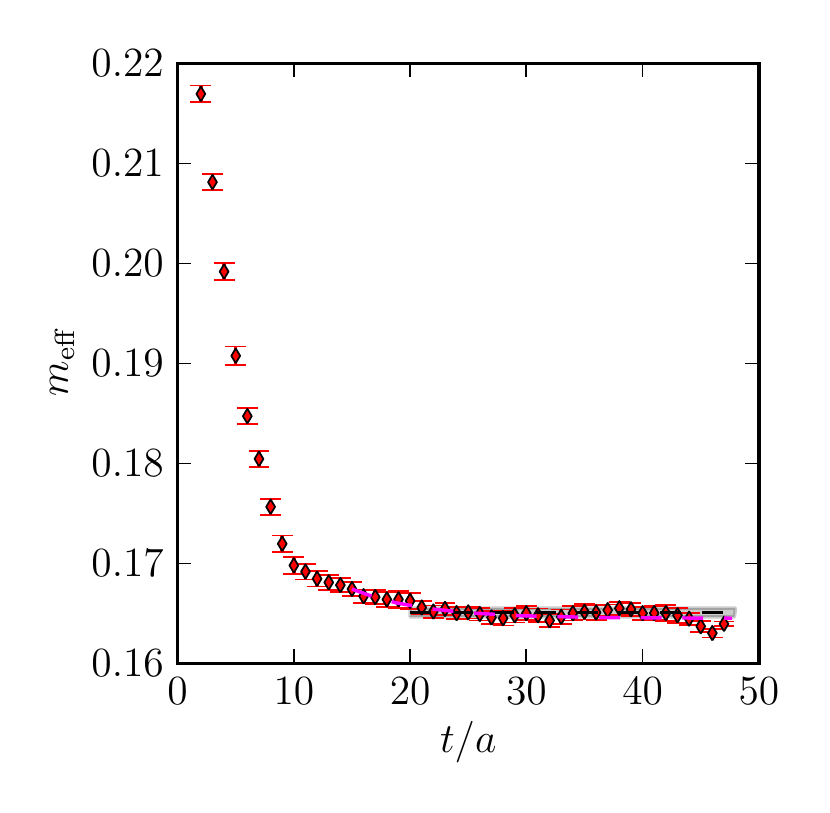}
\includegraphics[width=0.32\textwidth,keepaspectratio=]{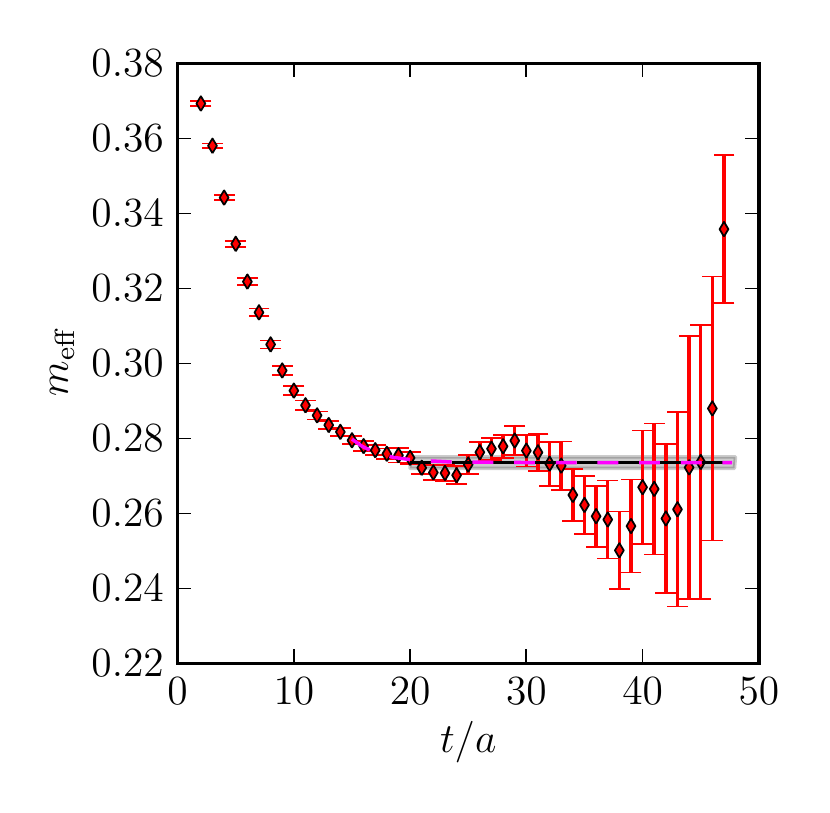}
\includegraphics[width=0.32\textwidth,keepaspectratio=]{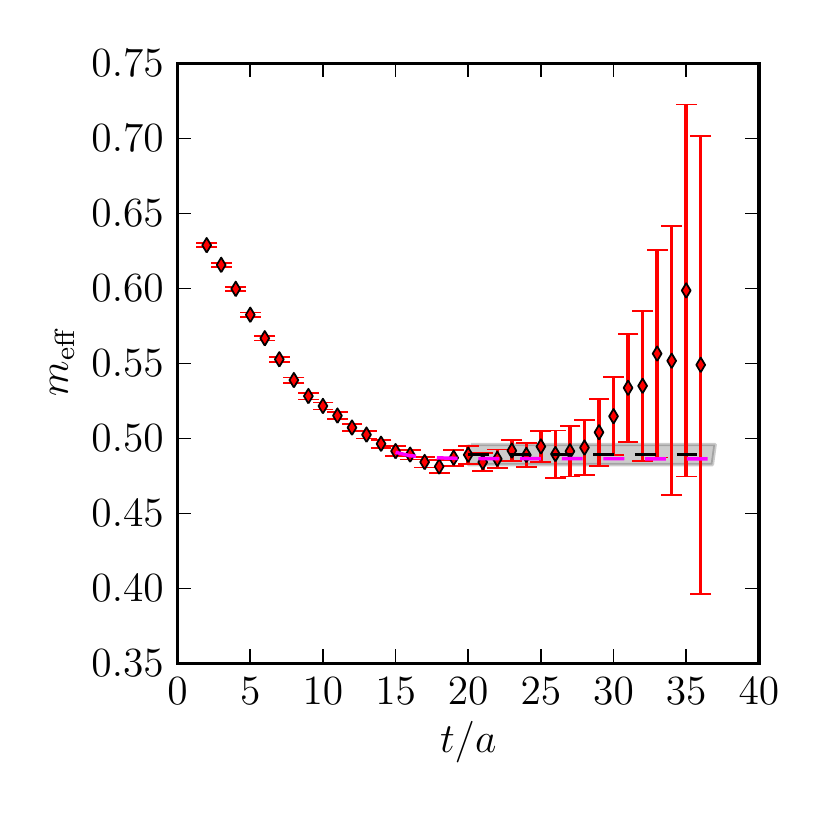}
\end{center}
\caption{Examples of effective mass plots with plateaux and two-state fits shown
         for a pseudoscalar and vector meson and a decuplet baryon channel,
         corresponding to a trial $K$, $K^*$ and $\Omega$ on the $N4$ ensemble,
         respectively.}
\label{fig:plateaux}
\end{figure}

From each measured correlator we extract the effective masses and apply
the following methods to extract estimates for the mass of the ground
state: First, a naive plateau fit is performed for each correlator,
starting at a reasonably large Euclidean time; secondly, taking the plateau value
as the starting point, a two-state fit including the leading excited state
contribution is performed with a larger time range.
Finally, another two-state fit with the same time range
is performed for the baryonic channel,
in which the gap between the ground and excited
state is fixed to the theoretically expected gap of $2m_\pi$ using the
measured pion mass.
Examples of effective mass plots are shown in fig.
\ref{fig:plateaux}.
\subsection{Fixing \ks}

In order to establish the physical value of the $\Omega$ baryon, we need to
fix the strange quark mass to its physical value. Since we intend to set the
scale from the $\Omega$ mass, a scale-free renormalization condition is needed
for this purpose, and we therefore will fix \ks\ by demanding that a ratio of
masses equals its physical value. Different choices of the ratio used correspond
to different renormalization conditions, which will introduce an ambiguity in the
scale through the resulting ambiguity in \ks.

\begin{figure}
\begin{center}
\includegraphics[width=0.6\textwidth,keepaspectratio=]{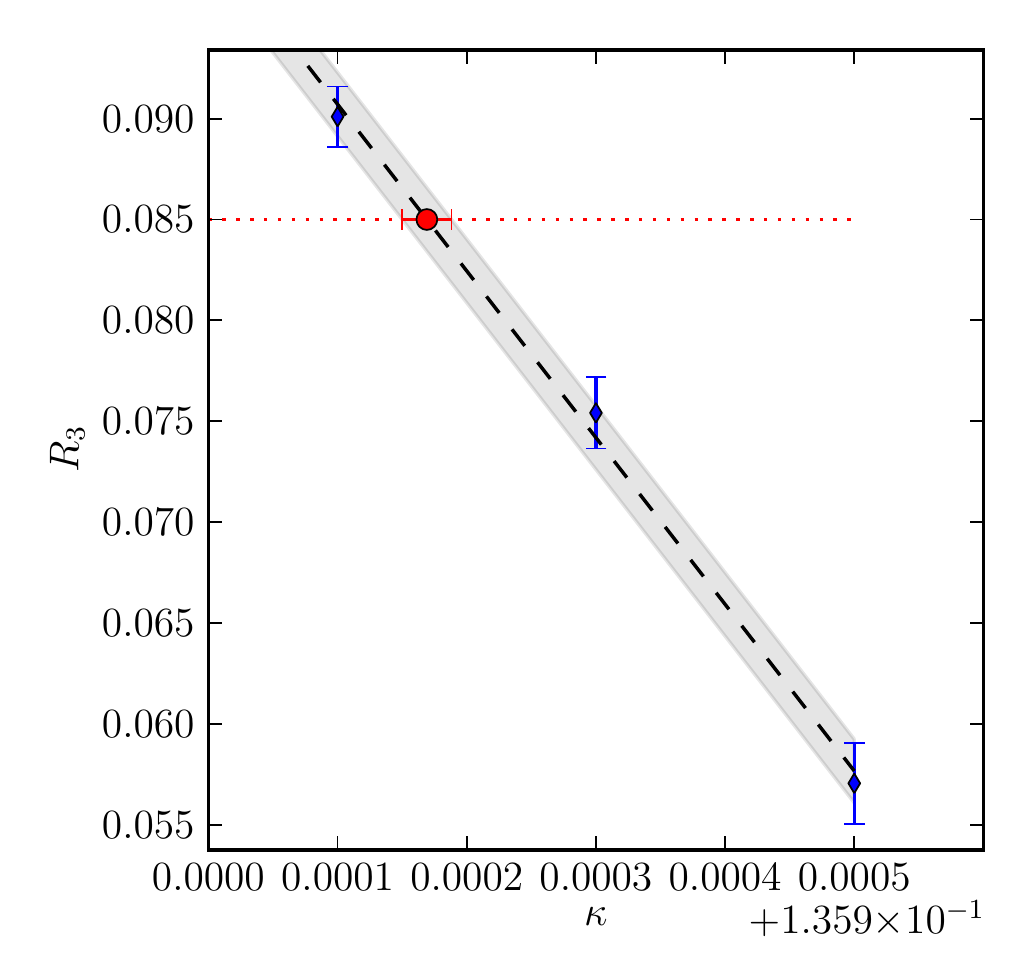}
\end{center}
\caption{An example of the interpolation to the physical value of the strange
         quark mass. The measured ratios are shown as blue diamonds,
         and the interpolated value as the red circle.}
\label{fig:interpol}
\end{figure}

The three ratios which we have chosen to fix \ks\ are
\begin{itemize}
\item $R_1 = m_K/m_{K^*}$, which is the ratio originally used by the CERN
      group in
      \cite{DelDebbio:2006cn},
\item a chirally improved variant $R_2 = (m_K^2-\frac{1}{2}m_\pi^2)/m_{K^*}^2$,
      from which the leading $m_q$ dependence of $R_1$ has been removed, and
\item $R_3 = (m_K^2-\frac{1}{2}m_\pi^2)/m_\Omega^2$, which is obtained by
      replacing the $K^*$ resonance by the stable $\Omega$ baryon in $R_2$.
\end{itemize}
Determining all three ratios $R_i$ requires us to measure $m_\pi$,
$m^{\rm qs}_{\rm PS}$, $m^{\rm qs}_{\rm V}$ and $m^{\rm sss}_{\bf 10}$
for a range of trial \ks\ at each value of $\beta$ and $\kappa_{\rm sea}$;
we then determine the value of \ks\ corresponding to the physical strange
quark mass by interpolating to the physical value of $R_i$. An example is shown
in fig.
\ref{fig:interpol}.

The difference in the value of \ks\ that results from the different mass fits
is generally less than the statistical errors, which implies that the systematic
errors from excited states are well under control. Similarly, the difference
between a linear and a quadratic interpolation in \ks\ or in $\ks^{-1}$ is
much smaller than the statistical errors. All statistical errors were determined
using the UWerr procedure
\cite{Wolff:2003sm}
for error propagation including the effects of autocorrelations.
In the following, we have chosen the results from the fit with a fixed gap 
of $2m_\pi$ as our final value.

\begin{figure}
\begin{center}
\includegraphics[width=0.33\textwidth,keepaspectratio=]{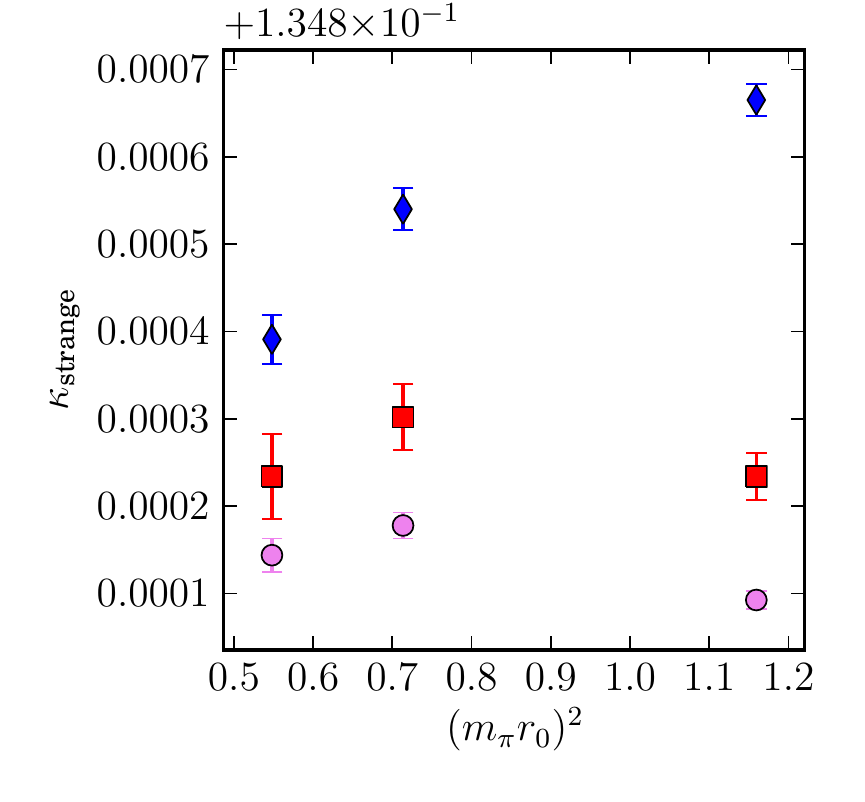}
\hskip-1em
\includegraphics[width=0.33\textwidth,keepaspectratio=]{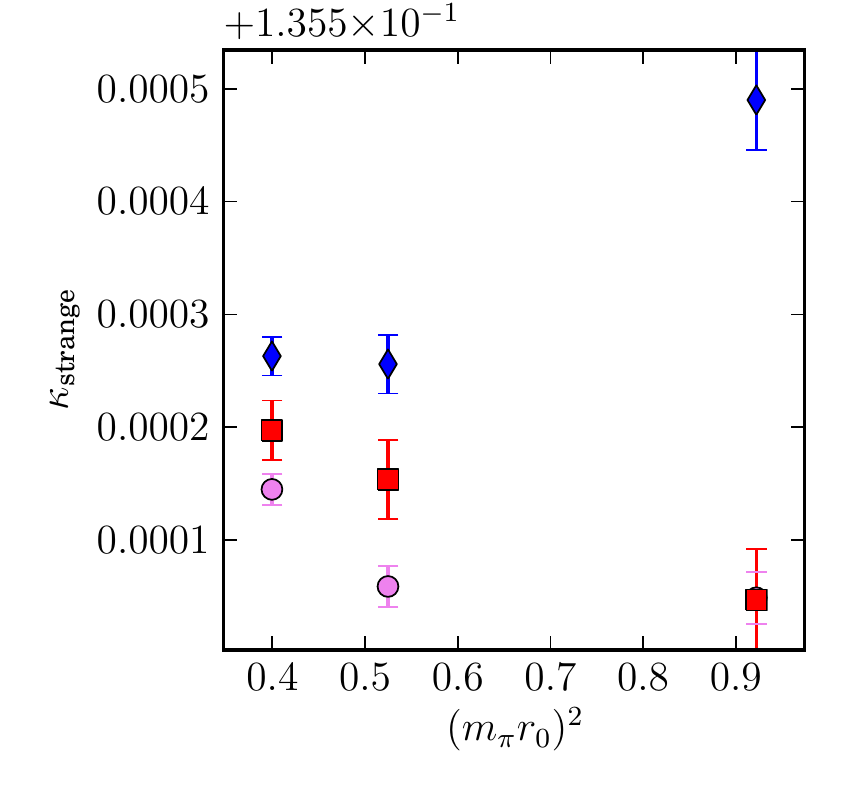}
\hskip-0.5em
\includegraphics[width=0.33\textwidth,keepaspectratio=]{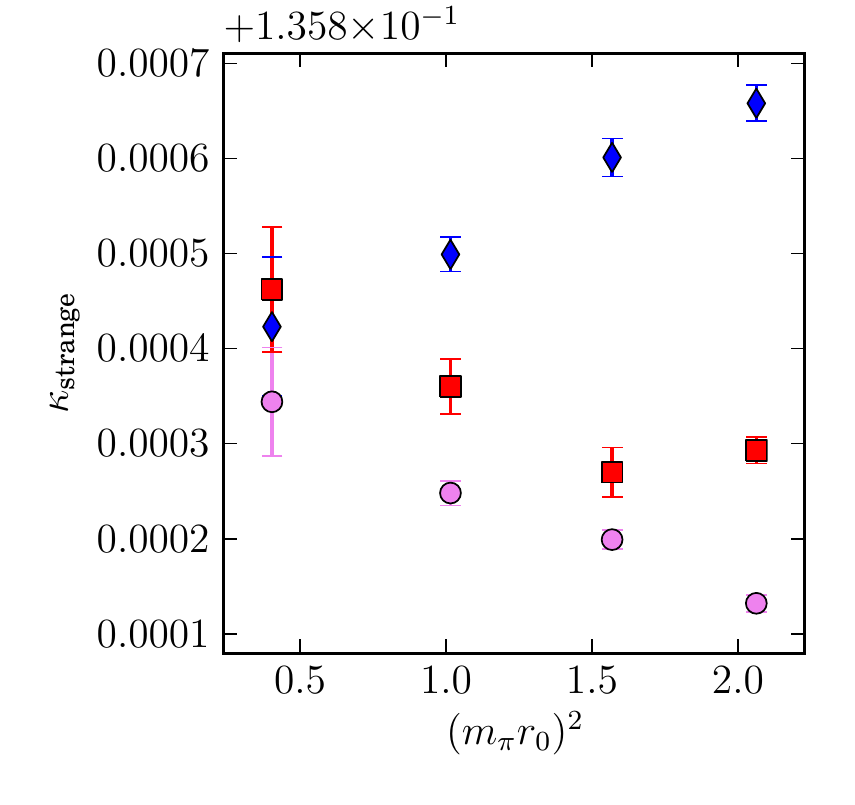}
\end{center}
\caption{The values of \ks\ from the three renormalization conditions studied.
         From left to right, the $\beta=5.2$, $5.3$ and $5.5$ ensembles are
         shown with the values obtained from $R_1$, $R_2$ and $R_3$
         plotted as blue diamonds, pink circles and red squares, respectively.}
\label{fig:kappachiral}
\end{figure}

On the other hand, the ambiguity in \ks\ from the choice of the renormalization
condition is significant, but decreases as expected towards the chiral limit,
as can be seen in fig.
\ref{fig:kappachiral}.
We find that condition 3, which employs the $\Omega$ baryon and a chirally
improved estimate of the strange quark mass, is the most stable and therefore
we choose it for our final analysis.

\subsection{Fixing the scale}

The value $r_0/a$ of the Sommer scale in lattice units has been determined
on the CLS configurations by Knechtli and Leder
\cite{Leder:LAT11}.
On each ensemble, we combine their results for $r_0/a$ and our results
for $am_\Omega$ using the UWerr procedure
\cite{Wolff:2003sm}
in order to take into account both the correlations between observables
and the autocorrelations within each ensemble (for an in-depth discussion
of the issues associated with the latter cf.
\cite{Schaefer:2010hu}).

Since our respective measurements for the $\beta=5.5$ ensembles were performed
on non-overlapping subsets, we combine the results on these ensembles neglecting
the correlations between $r_0$ and $m_\Omega$;
since we find that the error of $r_0 m_\Omega$ is dominated
by the error on $am_\Omega$, any resulting systematic error may be assumed
to be small.

\begin{figure}
\begin{center}
\includegraphics[width=0.48\textwidth,keepaspectratio=]{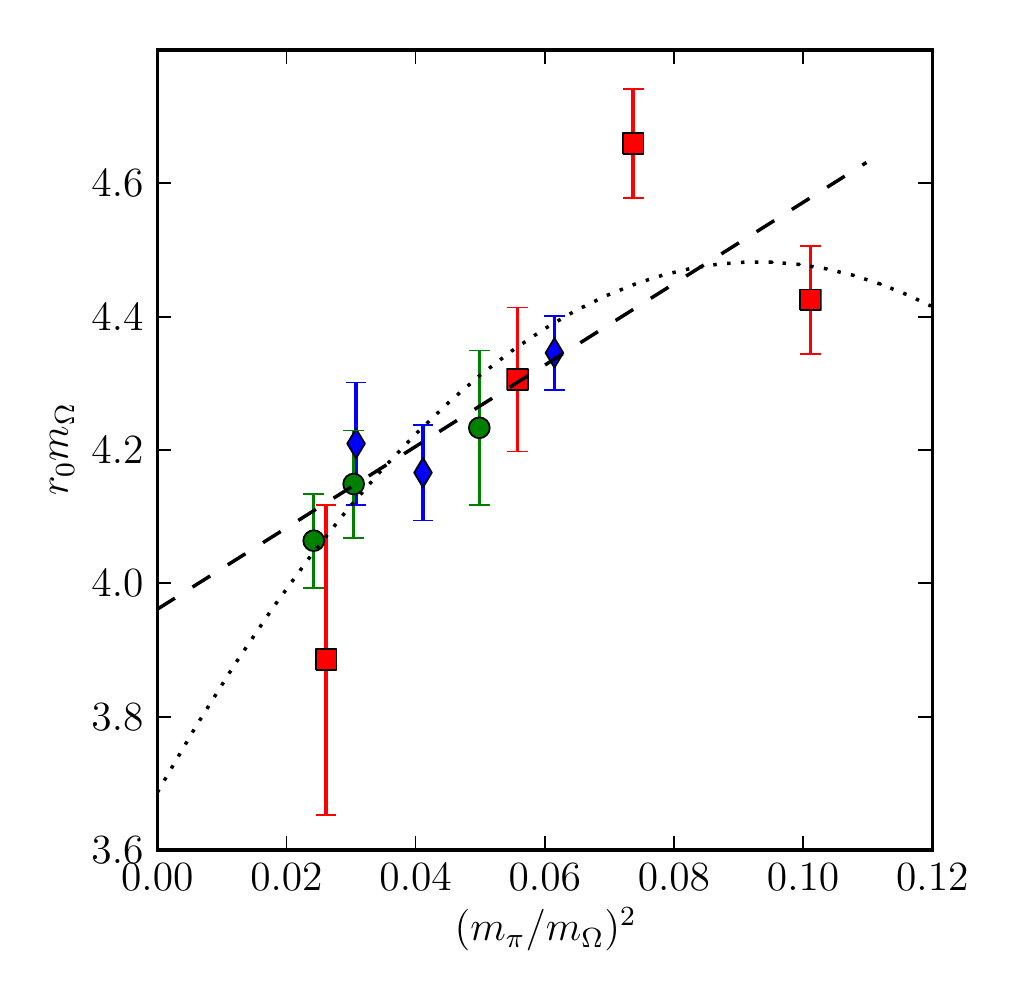}
\includegraphics[width=0.48\textwidth,keepaspectratio=]{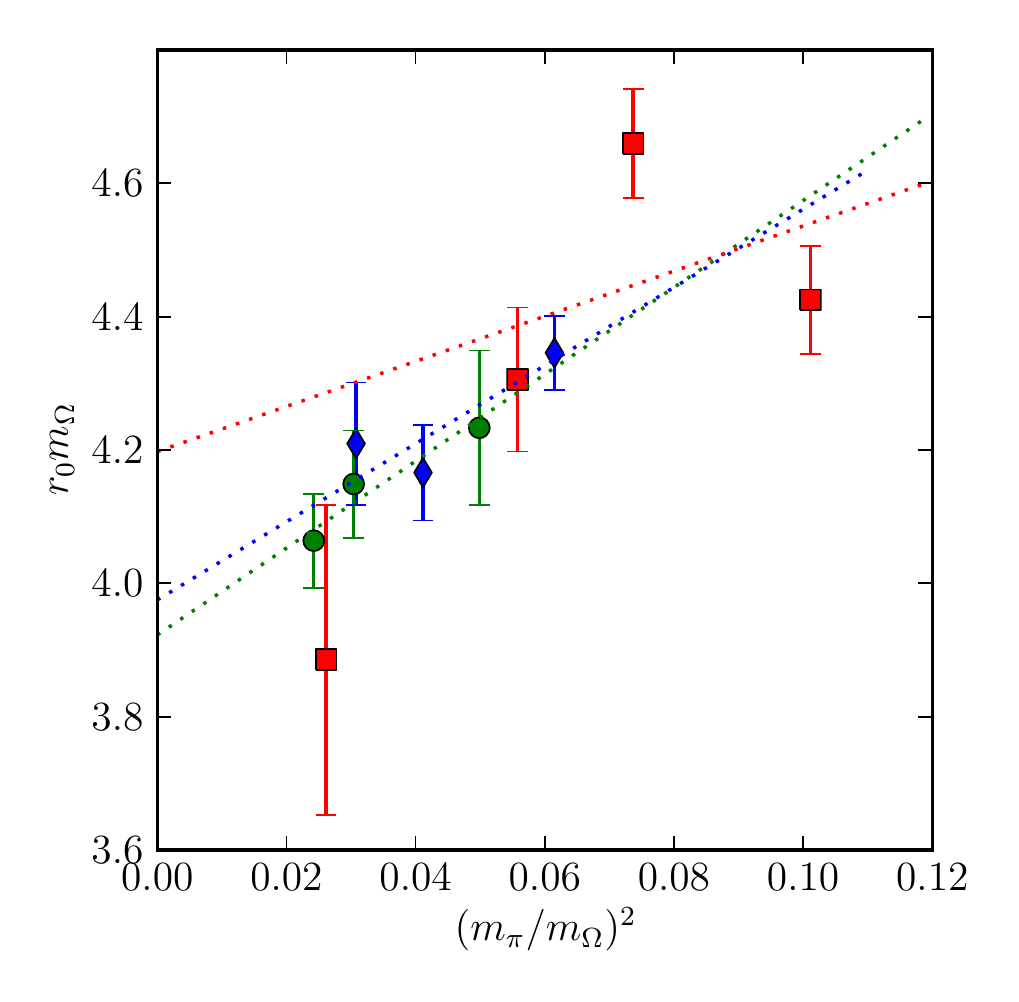}
\end{center}
\caption{Left: the joint linear and quadratic chiral extrapolations of
         $r_0 m_\Omega$ ignoring discretization effects; right: the separate
         chiral extrapolations of $r_0 m_\Omega$ at each value of $\beta$.
         The $\beta=5.2$, $5.3$ and $5.5$ ensembles are shown as blue diamonds,
         green circles and red squares, respectively.}
\label{fig:r0chiral}
\end{figure}

The finite combination $r_0 m_\Omega$ can now be extrapolated to the physical
point in $a^2$ and in $(m_\pi/m_\Omega)^2$. To estimate systematic uncertainties,
we used four different fits:
\begin{itemize}
\item a linear fit in $(m_\pi/m_\Omega)^2$, assuming no lattice artifacts,
\item a quadratic polynomial in $(m_\pi/m_\Omega)^2$, assuming no lattice artifacts,
\item a linear fit in $(m_\pi/m_\Omega)^2$ at each $\beta$ separately, followed by a linear fit of the chirally extrapolated results in $(a/r_0)^2$, and
\item a combined continuum and chiral extrapolation by a fit linear in both $(m_\pi/m_\Omega)^2$ and $(a/r_0)^2$.
\end{itemize}
The first two fits, which are made under the assumption that lattice artifacts
are absent from the combination $r_0 m_\Omega$,
are motivated by the scaling behaviour seen in
fig. \ref{fig:r0chiral}.
All fits agree within their respective errors.

\section{Results}

A combined continuum and chiral extrapolation of $r_0 m_\Omega$ gives
a physical value of $r_0 m_\Omega=3.99(12)(9)$, where the errors are
statistical and systematic, with the systematic error estimated from the
spread of the results obtained using the different fitting strategies
both for the mass extraction and in the extrapolation.

Combining this with the experimental value $m_\Omega=1672.45(29)$ MeV
\cite{PDG}
gives a value of $r_0=0.471(14)(10)$ fm for the Sommer scale, which is
compatible with the ETMC result
\cite{Alexandrou:2009qu}
of $r_0=0.465(6)(15)$ using the nucleon.

Performing a separate chiral extrapolation of $am_\Omega$ to the physical
pion mass at each value of $\beta$, we find the lattice spacings
at the physical point to be
\begin{center}
\begin{tabular}{llll}\hline\hline
$\beta$ & 5.2 & 5.3 & 5.5 \\\hline
$a$ [fm] & 0.079(3)(2) & 0.063(2)(2) & 0.050(2)(2)\\\hline\hline
\end{tabular}
\end{center}

We note that the lattice spacings obtained from the $\Omega$ baryon mass
are significantly finer than those previously obtained
\cite{Capitani:2009tg}
for the same ensembles using the CERN method
\cite{DelDebbio:2006cn}.
This is not entirely unexpected, since the CERN method uses
an unphysical reference point of $m_\pi/m_K=0.85$ to make contact with
the physical hadron masses, and was intended primarily for determining
the relative scale between different $\beta$ values rather
than an absolute value of the lattice scale.

The new values presented here are, however, compatible with the results
obtained in an entirely separate approach from $f_K$
\cite{Marinkovic:LAT11},
as well as with the results from an independent analysis
of the same data using a different error estimator and performing the
chiral extrapolation in $(m_\pi/m_N)^2$
\cite{Knippschild:PhD},
which gives us confidence in the robustness of the results.

\section{Conclusions}

The mass of the $\Omega$ baryon provides a way to set the scale in a way that
is particularly insensitive to the sea quark mass, and independent of
renormalization constants and effective theories. We find that the relatively
large statistical errors of the $\Omega$ baryon mass can be controlled, and
that the systematic errors due to excited states are also well under control;
hence the $\Omega$ provides an attractive means to set the scale.

Using $m_\Omega$, we have determined the Sommer scale to be
$r_0=0.471(14)(10)$ fm and have set the lattice spacing on the $N_f=2$ CLS
lattices. The lattice spacings determined in this way are compatible with
those found in an independent analysis employing $f_K$ to set the scale.
We find that the lattices are finer than had been previously expected.
What influence this will have on the question of finite-volume effects on
studies of hadronic structure
\cite{Wittig:SFB}
remains to be seen.

In the immediate future, we intend to increase the number of measurements on
the O7 ensemble, which will reduce the statistical error on the lattice spacing
at $\beta=5.5$; in the longer run, we will also include in our analysis
the G8 and N6 ensembles currently being generated, which should help us to
improve our control of the chiral extrapolation.

%%%

%%%

\end{document}